\documentclass[aps,prx,twocolumn,footinbib,floatfix,superscriptaddress]{revtex4-2}
\usepackage{graphicx}
\usepackage{indentfirst}
\usepackage{braket}
\usepackage{float}
\usepackage{amsmath}
\usepackage{physics}
\usepackage{amssymb}
\usepackage{CJK}
\usepackage{esint}
\usepackage{color}
\usepackage[T1]{fontenc}
\usepackage{subfigure}
\usepackage{amsfonts}
\usepackage{footmisc}
\usepackage{scrextend}
\usepackage{multirow}
\usepackage[outdir=./]{epstopdf}

\usepackage[english]{babel}
\usepackage{url}
\usepackage{comment}
\usepackage{array}
\usepackage{subfiles}
\usepackage{tikz}
\usetikzlibrary{shapes,arrows}
\usepackage{mathrsfs}
\usepackage[mathscr]{eucal}

\usepackage[hyperfootnotes=false]{hyperref}
\usepackage{cleveref}
\definecolor{darkblue}{rgb}{0,0,0.5}
\hypersetup{
colorlinks=true,
linkcolor=black,
filecolor=blue,
citecolor=darkblue,  
urlcolor=black,
}

\def\be{\begin{equation}}
\def\ee{\end{equation}}
\def\ba{\begin{eqnarray}}
\def\ea{\end{eqnarray}}
\def\bal{\begin{equation}\begin{aligned}}
\def\eal{\end{aligned}\end{equation}}

\def\bp{\begin{pmatrix}}
\def\ep{\end{pmatrix}}

\usepackage{bm}

\newcommand{\calN}{{\cal N}}

\newcommand{\1}{^{(1)}}

\newcommand{\QZ}[1]{{{\textcolor{black}{#1}}}}
\newcommand{\bz}[1]{{{\textcolor{black}{#1}}}}

\usepackage[normalem]{ulem}

\begin{document}

\title{Enhancing distributed sensing with imperfect error correction}% Force line 

%\title{Performance of finite squeezed GKP-TMS concatenation-codes in a heterogeneous environment} %another option, more general; AJB

\author{Boyu Zhou}
\affiliation{
Department of Physics, University of Arizona, Tucson, Arizona 85721, USA
}
\affiliation{
Department of Electrical and Computer Engineering, University of Arizona, Tucson, Arizona 85721, USA
}
\author{Anthony J. Brady}
\affiliation{
Department of Electrical and Computer Engineering, University of Arizona, Tucson, Arizona 85721, USA
}
\author{Quntao Zhuang}%
\email{zhuangquntao@email.arizona.edu}
\affiliation{
Department of Electrical and Computer Engineering, University of Arizona, Tucson, Arizona 85721, USA
}
\affiliation{
James C. Wyant College of Optical Sciences, University of Arizona, Tucson, Arizona 85721, USA
}

\date{\today}% It is always \today, today,
             %  but any date may be explicitly specified

\begin{abstract}

Entanglement has shown promise in enhancing information processing tasks in a sensor network, via distributed quantum sensing protocols. As noise is ubiquitous in sensor networks, error correction schemes based on Gottesman, Kitaev and Preskill (GKP) states are required to enhance the performance, as shown in [New J. Phys. {\bf 22}, 022001 (2020)] assuming homogeneous noise among sensors and perfect GKP states.
Here, we extend the analyses of performance enhancement to finite squeezed GKP states in a heterogeneous noise model. To begin with, we study different concatenation schemes of GKP-two-mode-squeezing codes. While traditional sequential concatenation schemes in previous works do improve the suppression of noise, we propose a balanced concatenation scheme that outperforms the sequential scheme in presence of finite GKP squeezing. We then apply these results to two specific tasks in distributed quantum sensing---parameter estimation and hypothesis testing---to understand the trade-off between imperfect squeezing and performance. For the former task, we consider an energy-constrained scenario and provide an optimal way to distribute the energy of the finite squeezed GKP states among the sensors. For the latter task, we show that the error probability can still be drastically lowered via concatenation of realistic finite squeezed GKP codes.

\end{abstract}

\maketitle

%\tableofcontents

\section{\label{sec:level1} Introduction}
%quantum sensing

Quantum sensing~\cite{giovannetti2006quantum, giovannetti2011advances,degen2017quantum} utilizes quantum phenomena to enhance measurement precision. The major benefit of quantum sensing, as opposed to classical sensing, is the ability of quantum strategies to provably outperform any classical strategy~\cite{pirandola2018advances}, which has resulted in a growth of the application spaces like, e.g., the laser interferometer gravitational-wave observatory~\cite{abadie2011gravitational,abbott2016observation}, quantum illumination~\cite{lloyd2008enhanced,tan2008quantum,zhuang2017optimum,zhuang2021ultimate}, quantum reading~\cite{pirandola2011quantum,zhuang2020entanglement,ortolano2021}, and distributed quantum sensing (DQS)~\cite{zhang2021distributed}.

DQS, in particular, is an intriguing category of its own, as it endeavors to estimate a global parameter of a system, such as a weighted average of displacements~\cite{zhuang2018distributed,xia2020demonstration} or phase rotations~\cite{guo2020distributed} -- as opposed to estimating a set of unknown (local) parameters or a single (local) parameter. Recent theoretical works \cite{proctor2018multiparameter,ge2017distributed,zhuang2018distributed,eldredge2018optimal}, have shown that DQS protocols can boost the performance of estimating such global features by leveraging an entangled set of sensors to probe the system of interest. One can achieve this enhanced performance, for example, with discrete variable (DV) quantum sensors~\cite{ge2017distributed,eldredge2018optimal,zhao2021field,liu2021distributed} (e.g. qubits) or continuous variable (CV) quantum sensors~\cite{zhuang2018distributed} (e.g. modes of the electromagnetic field).
DQS also enables further applications, such as the supervised learning assisted by an entangled sensor
network (SLAEN)~\cite{zhuang2019physical,xia2021quantum}, which provides a new route in achieving quantum advantage in machine learning~\cite{biamonte2017quantum,havlivcek2019supervised,rebentrost2014quantum}. 

The performance enhancement in DQS protocols decays in the presence of loss and noise due to the fragility of quantum systems. On the other hand, quantum error correction (QEC)~\cite{shor1995scheme,gottesman2001encoding} has been developed to protect quantum information against noise and error. In recent years, we have seen great progress in experiments~\cite{grimm2020stabilization,campagne2020quantum}.
However, these qubit-into-an-oscillator schemes are designed to protect discrete variables, and we cannot apply these schemes directly to CV quantum information processing tasks. To protect quantum information in CV systems, Noh et al. recently developed a GKP-two-mode-squeezing (GKP-TMS) code~\cite{noh2020encoding}, which is a method of encoding a single mode CV system (an oscillator) into multiple oscillators, utilizing entangling Gaussian operations on GKP ancilla.
%error correction code, Liang Jiang, Jing Wu
Previous research \cite{zhuang2020distributed,wu2021continuous} has showed the GKP codes in its ideal implementation can improve the DQS protocols. 

In this paper, we further analyze the finite squeezed GKP-TMS code in a more generalized scenario where the Gaussian noises are heterogeneous. 
Since successively concatenating a QEC code generally suppresses noise with each level of concatenation, we begin our analyses with the design of concatenation. In the presence of the GKP finite squeezing noise, we show that the sequential concatenation proposed in previous works fails to achieve the minimum noise; instead, we propose a balanced concatenation scheme to approach the ultimate limit set by the finite squeezed GKP states. Then we proceed to design a general DQS protocol with noisy GKP-TMS codes. In a photon number constrained scenario, we also provide an optimal photon distribution of noisy GKP states to minimize the variance, which fully depends on the weights of parameters in the DQS protocol. 
Two specific applications, parameter estimation in DQS and hypothesis testing in SLAEN, are researched to illustrate the performance of DQS with finite squeezed GKP-TMS codes.

\section{Overview}

\begin{figure}
    \centering
    \includegraphics[width=\linewidth]{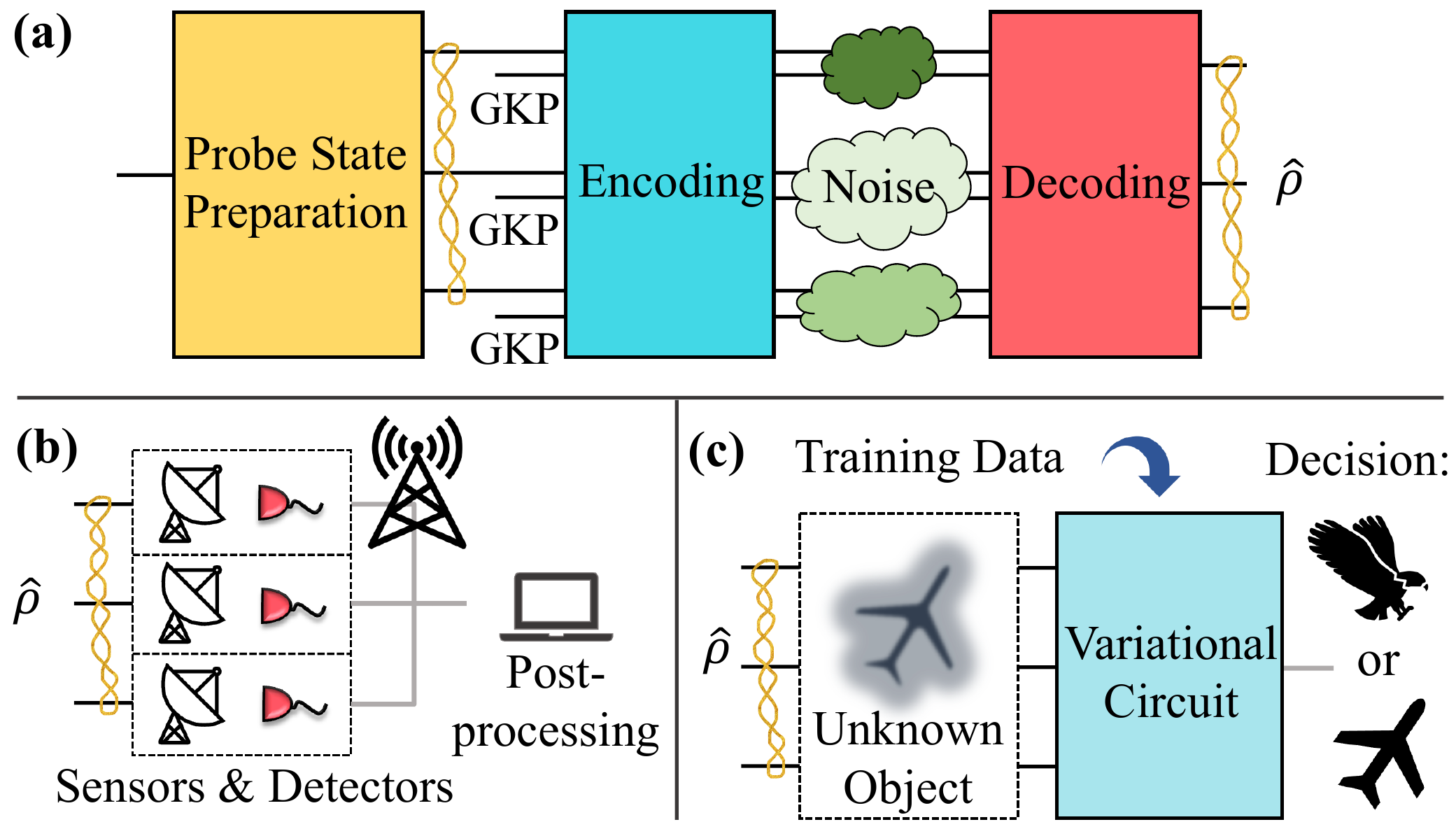}
    \caption{(a) CV-QEC scheme to protect an entangled probe state from the (heterogeneous) noise accumulated during the distribution of sensor probes. The state $\hat{{\rho}}$ is the approximate entangled probe state after QEC. One can use the approximate probe state $\hat{{\rho}}$ (b) as an input to a DQS protocol \cite{zhuang2018distributed,zhang2021distributed} or (c) as an input for SLAEN \cite{zhuang2019physical}.}
    \label{fig:0}
\end{figure}

In this paper, we consider and analyze the performance of the general setup depicted in Fig.~\ref{fig:0} (a), which we heuristically describe here. We prepare an entangled probe state across a set of modes (for example, by passing a single-mode squeezed vacuum through a balanced beam-splitter network) and then encode the probe-state via a GKP-TMS code with finite-squeezed GKP ancilla states. The GKP-TMS code protects the entangled probe against potential noise accumulated during the distribution of the modes to receiver nodes across noisy quantum channels (depicted by clouds in the figure). A joint decoding-strategy results in an output entangled-state ($\hat{{\rho}}$ in the figure) that approximates the initial probe state. Once the decoding has been successfully performed, the receiver nodes can use the recovered entangled-probe state i) as a resource for an entanglement-enhanced DQS protocol (see Fig.~\ref{fig:0}(b) and Ref.~\cite{zhuang2020distributed}), or ii) as a resource to enhance the performance of machine learning tasks, such as data classification for target recognition (see Fig.~\ref{fig:0}(c) and Ref.~\cite{zhuang2019physical}) which relies on input data acquired from a set of distributed sensors. As we describe in detail throughout the rest of this paper, protecting valuable quantum probe-states with GKP-TMS codes can significantly boost the performance of quantum strategies as applied to these various tasks, especially when there is a significant amount of noise in the distribution process. We provide a more technical breakdown of the paper below.

In Sec.~\ref{III} A, for the sake of pedagogy and completeness, we introduce the finite squeezed GKP-TMS code in a generalized setting where the Gaussian noises are heterogeneous. We also derive asymptotic expressions to show adverse effects of finite squeezing of approximate GKP states. In Sec.~\ref{III B}, to further reduce quadrature noise in our logical states, we analyze concatenation codes. We briefly discuss the basic types of concatenation schemes and their performance limitations, where we derive analytic lower bounds.

In Secs.~\ref{IV} and \ref{V}, we apply CV-QEC to two specific tasks of DQS and SLAEN. In Sec.~\ref{IV A}, we first show how the GKP-TMS QEC code can enhance the robustness of sensing protocols against imperfections. We assess the performance of these codes in Sec.~\ref{IV B}, where we consider protocol optimization under energy constraints. Finally, in Sec.~\ref{V}, we introduce a simple binary channel-discrimination task to demonstrate how GKP-TMS code can improve supervised learning tasks in a distributed-sensor setting.

\section{Analysis of finite squeezed GKP-TMS code}
\label{III}
\subsection{Introduction of finite squeezed GKP-TMS code}
\label{III A}

Additive Gaussian noise is ubiquitous in many realistic bosonic systems, however such noise cannot be corrected by using quantum error correction codes with only Gaussian resources. This is due to the no-go theorem for Gaussian QEC schemes \cite{eisert2002distilling,niset2009no}, which states that, to correct Gaussian errors, non-Gaussian resources are required \cite{zhuang2018resource,takagi2018convex}. Ref.~\cite{noh2020encoding} demonstrated such a non-Gaussian quantum error correction scheme that uses GKP states to encode an oscillator into many oscillators in a robust way, thus bypassing the no-go theorem for Gaussian codes.

One attractive characteristic of a GKP state is that it can, in some sense, circumvent the Heisenberg uncertainty principle. According to the Heisenberg uncertainty principle, the position operator $\hat{q}\equiv (\hat{a}+\hat{a}^{\dagger})/\sqrt{2}$ and momentum operator $\hat{p}\equiv i(\hat{a}^{\dagger}-\hat{a})/\sqrt{2}$ cannot be accurately and simultaneously measured. Here $\hat{a}$ is the annihilation operator of the field. The canonical GKP state (or the grid state) \cite{gottesman2001encoding} is defined as the simultaneous eigenstates of the two commuting displacement operators $\hat{S}_q \equiv e^{i\sqrt{2\pi}\hat{q}}$ and $\hat{S}_p \equiv e^{-i\sqrt{2\pi}\hat{p}}$, and $\left [\hat{S}_p,\hat{S}_q \right ]=0$. Since these two displacement operators commute with each other, we can therefore measure these unitary operators simultaneously--- thereby also measuring $\hat{q}$ and $\hat{p}$ simultaneously, up to a modulo $\sqrt{2\pi}$ ambiguity. Though, this only holds for a measurement scheme assisted by the unrealistic ideal GKP state, which is a superposition of infinitely many, infinitely squeezed states. We can however utilize a realistic, approximate (or 
imperfect) GKP state with finite squeezing. Formally, this state can be written as
\begin{equation}
\begin{aligned}
\ket{\mathrm{\rm GKP}_{\Delta}}  &\propto \sum_{t=-\infty}^{\infty}e^{-\pi \Delta^2 t^2} \int e^{-(q-\sqrt{2\pi }t)^2/2\Delta^2}\ket{q}\mathrm{d}q
\\& \propto \sum_{t=-\infty}^{\infty}e^{-\Delta^2 p^2 /2} \int e^{-(p-\sqrt{2\pi }t)^2/2\Delta^2}\ket{p}\mathrm{d}p,
\end{aligned}
\end{equation}
where $t\in \mathbf{Z}$ takes all integer values.
Observe that the wave function has a series of Gaussian peaks, each with width $\Delta$. As $\Delta \rightarrow 0$, the finite squeezed GKP state approaches the ideal GKP state. The mean photon number of the GKP state is given by~\cite{vuillotquantum} $\bar{n}=\langle a^{\dagger}a \rangle = \langle p^2+q^2 \rangle/2-1/2 \approx 1/({2\Delta^2})$.

%Another way to obtain the approximate GKP state is to apply a Gaussian envelope operator $\mathrm{exp}[-\Delta \hat{n}]$ to the ideal GKP state \cite{terhal2016encoding},
%\begin{equation}
 %   \ket{\mathrm{\rm GKP}_{\Delta}}  \propto e^{-\Delta \hat{n}} \ket{\mathrm{\rm GKP}}.
%\end{equation}

We now briefly introduce the GKP-TMS code \cite{noh2020encoding} with finite squeezed GKP states. An arbitrary state $\ket{\psi}$ in the first bosonic mode (the data mode) can be encoded as
\begin{equation}
\ket{\psi_{\rm L}}=\hat{\mathrm{TS}}(G) \ket{\psi}\ket{\mathrm{\rm GKP}_\Delta},
\end{equation}
where $\ket{\mathrm{\rm GKP}_\Delta}$ is the finite squeezed GKP state with width $\Delta$ in the second mode (the ancilla mode), and $\hat{\mathrm{TS}}(G)$ is the two-mode squeezing operation with gain $G$. In what follows, we will consider how heterogeneous additive Gaussian noise errors (i.e. different noises acting on the data mode versus the ancilla mode) affect the properties of the logical output and thus the performance of the finite squeezed GKP-TMS code.

%In terms of the quadrature operators, a two-mode squeezing transformation corresponds to the following symplectic matrix,
%\begin{equation}
%\mathbf{S}_{\rm{TS}}=\begin{pmatrix} 
%\sqrt{G}\bm{I} & \sqrt{G-1}\bm{Z} \\ \sqrt{G-1}\bm{Z} & \sqrt{G}\bm{I} 
%\end{pmatrix},
%\end{equation}
%acting between the two modes, where $\bm{I}=\text{diag}(1,1)$ is the identity matrix, and $\bm{Z}=\text{diag}(1,-1)$ is the Pauli Z matrix. 

We need to supply two GKP states to implement the GKP-TMS code: one in the input of the ancilla mode and one more GKP state for quadrature measurements of the ancilla. \bz{The additional GKP state allows for a {non-destructive} quadrature measurement of the ancilla, but following this non-destructive measurement, a {destructive} quadrature measurement can be performed directly on the GKP ancilla, as the ancilla GKP state need not be preserved after the error correction round.} To model the finite squeezing effect, each GKP state is corrupted by an additive Gaussian noise channel $\mathcal{N}[\sigma_{\rm GKP}]$~\cite{noh2020encoding}, with $\sigma_{\rm GKP}^2=\Delta^2/2=1/(4\bar{n})$.
In the decoding procedure, we apply an inverse of the encoding circuit $\hat{\mathrm{TS}}^{\dagger}(G)$, which correlates the independent additive noises to produce the correlated errors $z_q^{(1)},z_p^{(1)}$ for the data mode and $z_q^{(2)},z_p^{(2)}$ for the ancilla mode. Then we measure the ancilla noise, $z_q^{(2)}$ and $z_p^{(2)}$ modulo $\sqrt{2\pi}$. The estimated results can be modeled as,
\begin{equation}
\begin{aligned}
& \tilde{z}_q^{(2)}=R_{\sqrt{2\pi}}(z_q^{(2)}+\xi_q^{\rm GKP}),
\\& \tilde{z}_p^{(2)}=R_{\sqrt{2\pi}}(z_p^{(2)}+\xi_p^{\rm GKP}),
\end{aligned}
\label{final_estimator}
\end{equation}
where independent and identically distributed (iid) zero-mean Gaussian random variables $\xi_q^{\rm GKP}$ and $\xi_p^{\rm GKP}$ are additional noise due to the two noisy GKP states and have variance $2\sigma^2_{\rm GKP}$. The function $R_x(z) \equiv z-n^*(z)x$ is a general module function, with $n^*(z)\equiv \mathrm{argmin}_{n\in \mathbf{Z}}\abs{z-nx}$. 
To remove the noise $z_q^{(1)}, z_p^{(1)}$ from the data mode, we can apply the counter displacement operations based on the measurement results. Similar to Ref.~\cite{noh2020encoding}, we take the Gaussian-approximated minimal-variance estimator of the noise, and after correction the noise in the output quadratures of the data mode is then

\bz{\begin{equation}
\begin{aligned}
& \xi_q^{\rm out} \equiv z_q^{(1)}+\frac{t}{s+2\sigma_{\rm GKP}^2}\tilde{z}_q^{(2)},
\\& \xi_p^{\rm out} \equiv z_p^{(1)}-\frac{t}{s+2\sigma_{\rm GKP}^2}\tilde{z}_p^{(2)},
\end{aligned}
\label{final_noise}
\end{equation}
where $t=\sqrt{G(G-1)}(\sigma_1^2+\sigma_2^2)$, and $s=(G-1)\sigma_1^2+G\sigma_2^2$} are the standard deviations for the \QZ{additive noise at} data mode and ancilla mode; \QZ{while $\sigma_1$ and $\sigma_2$ are the standard deviations of the original additive noise on the data and the ancilla mode before error correction.} We assume $\sigma_1\le \sigma_2$ as we can always choose the less noisy mode as the data mode to achieve a better performance~\cite{wu2021continuous}. Fig.~\ref{fig:0} (a) illustrates the general sketch of the scheme. %The probability density function can then be calculated and we can have the variance of the output logical quadrature noise.

Ref.~\cite{noh2020encoding} has analyzed the adverse effects of using approximate GKP states, and they have given the variance of the output logical quadrature noises, when the noise model is homogeneous across different modes. Using similar methods, from the final noise Eq.~\eqref{final_noise}, we can derive the quadrature variances, $\text{Var}[\xi_q^{\rm out} ]=\text{Var}[\xi_p^{\rm out} ]=\sigma_{\rm L}^2$ in a more realistic situation, where the additive Gaussian noise errors can be  heterogeneous across modes (see Appendix \ref{appendix A}),
\begin{align}
\sigma_{\rm L}^2&=\frac{\sigma_1^2 \sigma_2^2}{s}+\frac{2t \sigma_{\rm GKP}^2}{s+2\sigma_{\rm GKP}^2}
\nonumber
\\
&
+\sum_{n=1}^{n=\infty} \frac{2\pi t (2n-1)}{s+2\sigma_{\rm GKP}^2} \mathrm{Erfc}\left(\frac{(2n-1)\sqrt{\pi}}{2\sqrt{s+2\sigma_{\rm GKP}^2}}\right),
\label{eq:sigma_real_main}
\end{align}
where $t=\sqrt{G(G-1)}(\sigma_1^2+\sigma_2^2)$ and $\mathrm{Erfc}(x)=1-\left(2/\sqrt{\pi}\right)\int_0^z dt \exp\left(-t^2\right)$ is the complementary error function.  In general, one minimizes the noise variance by tuning the gain $G$.

To obtain insights, we perform asymptotic analyses by keeping the $n=1$ term in Eq.~\eqref{eq:sigma_real_main}. When $\sigma_1\approx\sigma_2$, we can obtain the optimum gain (see Appendix~\ref{appendix A} for details)
\begin{equation}
G\simeq \frac{1}{2}-\frac{\sigma_{\rm GKP}^2}{\bar{\sigma}^2}+\frac{\pi}{8\bar{\sigma}^2}\frac{1}{\ln{\left[{\pi^{\frac{3}{2}}}/{2(\bar{\sigma}^4-4\sigma_{\rm GKP}^4)}\right]}},
\label{eq:G}
\end{equation}
that minimizes the noise variance
\begin{equation}
\sigma_{\rm L}^2\simeq 2\sigma_{\rm GKP}^2+ \frac{4(\bar{\sigma}^4-4\sigma_{\rm GKP}^4)}{\pi}\ln{  \frac{\pi^{\frac{3}{2}}}{2(\bar{\sigma}^4-4\sigma_{\rm GKP}^4)} },
\label{eq:sigma_L}
\end{equation}
where $\bar{\sigma}=\sqrt{\sigma_1 \sigma_2}$ is the geometric mean of the input noises. \bz{Observe that this asymptotic expression corresponds to the non-trivial regime, $\sigma_{\rm GKP}<{\bar{\sigma}}/{\sqrt{2}}$, where the code is effective. For $\sigma_{\rm GKP}>{\bar{\sigma}}/{\sqrt{2}}$, the optimal gain is trivially $G=1$; i.e., it is best to not encode if the GKP ancilla is too noisy. These results are coincident with the numerics; see App.~\ref{appendix A} for more details. Since $\sigma_{\rm GKP}^2=\Delta^2/2=1/(4\bar{n})$, this condition thus sets a constraint on the least number of photons in the GKP states required for QEC (which also depends on the values of initial noises $\sigma_1$ and $\sigma_2$).} 

Although in this case ($\sigma_1\approx\sigma_2$) the asymptotic results are symmetric between $\sigma_1$ and $\sigma_2$, we note that in general Eq.~\eqref{eq:sigma_real_main} is not symmetric. When the GKP state is ideal ($\sigma_{\rm GKP}=0$), Eq.~\eqref{eq:G} and Eq.~\eqref{eq:sigma_L} degenerate to the results found in in Ref. \cite{noh2020encoding}. When $\sigma_1 \ll \sigma_2$, similar asymptotic results can be derived (see Appendix~\ref{appendix A})
\begin{align}
\sigma_{\rm L}^2\simeq 2\sigma_{\rm GKP}^2+ &\frac{4\left(\bar{\sigma}^4 -2\sigma_2^2 \sigma_{\rm GKP}^2 -4\sigma_{\rm GKP}^4\right)}{\pi}
\nonumber
\\
&\times\ln{  \frac{\pi^{\frac{3}{2}}}{2\left(\bar{\sigma}^4 -2\sigma_2^2 \sigma_{\rm GKP}^2 -4\sigma_{\rm GKP}^4\right)} }.
\label{sigma1small}
\end{align}

Apparently, $\sqrt{2}\sigma_{\rm GKP}<\sigma_1$ is required for error correction to be beneficial, as otherwise Eq.~\eqref{eq:sigma_L} and Eq.~\eqref{sigma1small} lead to $\sigma_{\rm L}>\sigma_1$. When this condition is satisfied, we also see that $\sigma_{\rm L}^2\ge 2\sigma_{\rm GKP}^2$ from the asymptotic formula, which can also be proven rigorously (see Appendix~\ref{appendix D} for a proof). Due to the extra noise added from the finite squeezed GKP states during encoding and decoding, the variance of the output noise $\sigma_{\rm L}^2$ cannot be decreased lower than $2\sigma_{\rm GKP}^2$. This is also one reason that the concatenation codes have this quantity as a lower bound, as we will discuss in the next section.

\subsection{Concatenation codes}
\label{III B}
\begin{figure}
    \centering
    \includegraphics[width=0.95\linewidth]{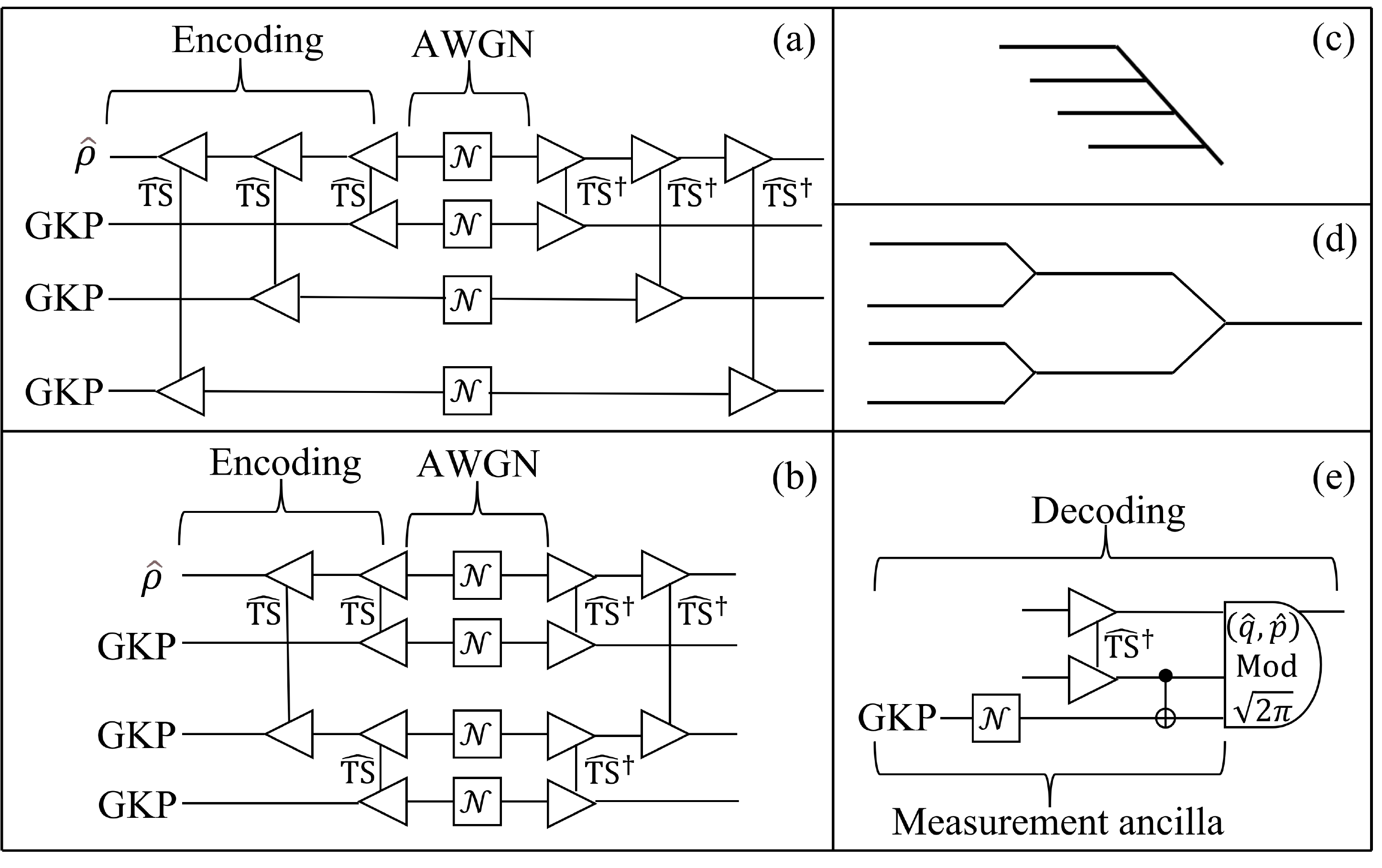}
    \caption{Different schemes for concatenation codes. (a) and (c) are circuit diagram and simplified schematic of a sequential scheme. (b) and (d) are circuit diagram and simplified schematic of a balanced scheme. For simplicity, we did not show the measurement of quadratures module $\sqrt{2\pi}$, which needs another GKP ancilla to be performed. (e) is the Decoding circuit, where we use the second noisy GKP state. `$\hat{\rm TS}$' represents two-mode squeezing and `$\hat{\rm TS}^\dagger$' represents the inverse two-mode squeezing. $\calN$ represents the additive white Gaussian noise (AWGN). }
    \label{fig.1}
\end{figure}

Concatenating QEC codes can further reduce the noise as compared to a single layer of error correction. In this section, we consider the different concatenation schemes based on the original finite squeezed GKP-TMS code. We evaluate the convergence rate and the limiting value of each concatenation scheme. Due to the non-Gaussian aspects of the probability density after a single layer of error correction, calculations can be cumbersome; guided by Ref.~\cite{wu2021continuous}'s results, we approximate the distribution as Gaussian and therefore only numerically keep track of the variance through the concatenations, which is a good approximation in the parameter region we care about.

As shown in Fig.~\ref{fig.1}, we consider the two basic concatenation schemes, (a) and (b). Scheme (a) is previously considered~\cite{noh2020encoding,wu2021continuous} and easy to understand: It just adds layers of the original circuit sequentially. We call this scheme `sequential'. 
%The ancillary, finite squeezed GKP states can be arbitrary, with the only restriction being that $\sigma_{\rm GKP}$ should be not too large that when $\sigma_{\rm GKP} \geq \sigma/\sqrt{2}$ that the error correction code is ineffective. 
To simplify the calculations in the proceeding, we assume that all ancilla GKP states are identical, with noise $\sigma_{\rm GKP} < \sigma/\sqrt{2}$. To simplify the comparison between the two concatenation schemes, we let the initial noise in each layer's ancillary mode to be identically $\sigma$.
For ideal GKP states, we can get an arbitrary output $\sigma_{\rm L}$ as we can successively add more concatenation layers, however, for finite squeezed GKP states, as Eq.~\eqref{eq:sigma_L} and Eq.~\eqref{sigma1small} show, successively adding layers will not lower $\sigma_{\rm L}$ below $\sqrt{2 }\sigma_{\rm GKP}$, which establishes an ultimate lower bound. 

\bz{To obtain the lower bound which is more close to the numerical result, we need to use  Eq.~\eqref{sigma2small} (see in App.~\ref{appendix A}),}
since $\sigma_1$, and $\sigma_2$ have large difference after many layers. For the $\ell$th layer, we have the logical output noise
\begin{equation}
\begin{aligned}
(\sigma_{\rm L}^{(\ell)})^2\simeq& 2\sigma_{\rm GKP}^2+ \frac{4\left((\bar{\sigma}_{\rm L}^{(\ell-1)})^2\sigma^2+2\sigma^2\sigma_{\rm GKP}^2-4\sigma_{\rm GKP}^4\right)}{\pi}
\\& \times \ln{  \frac{\pi^{\frac{3}{2}}}{2\left((\bar{\sigma}_{\rm L}^{(\ell-1)})^2\sigma^2+2\sigma^2\sigma_{\rm GKP}^2-4\sigma_{\rm GKP}^4\right)} },
\label{eq:sigma^n_sequential}
\end{aligned}
\end{equation}
where $\sigma_{\rm L}^{(\ell-1)}$ is the logical output noise of the $(\ell-1)$th concatenation layer, and  $\bar{\sigma}_{\rm L}^{(\ell-1)}=\sqrt{\sigma_{\rm L}^{(\ell-1)} \sigma}$. As the number of concatenation layers increase, the noise will eventually converge. Again, we note that the variance is always lower bounded by $2\sigma_{\rm GKP}^2$. The converged value $\sigma_{\infty}$ can be numerically obtained via solving Eq.~\eqref{eq:sigma^n_sequential} after setting $\bar{\sigma}_{\rm L}^{(\ell)}=\bar{\sigma}_{\rm L}^{(\ell-1)}=\sigma_{\infty}$. %It serves as a lower bound of the noise in a sequential scheme.

%The asymptotic result of sequential type is:
%\begin{equation}
%\begin{aligned}
%2\sigma_{\rm GKP}^2 + 4\frac{\sigma_{\rm GKP}^4+\sigma^2\sigma_{\rm GKP}^2}{\sigma^2} -\frac{\pi\left(\sigma_{\rm GKP}^4-\sigma^2\sigma_{\rm GKP}^2 \right)}{\sigma^4 W_0\left(-\frac{2\sigma_{\rm GKP}^4e^{\frac{\pi}{4\sigma^2}}}{\sqrt{\pi}\sigma^2}\right)}
%\label{eq:lb for sequential}
%\end{aligned}
%\end{equation}
%where $W_0$ is Lambert W function (or product logarithm), and the proof can be seen in Appendix~\ref{appendix E}.

For the scheme in Fig.~\ref{fig.1} (b), each layer is a repetition of the last layer's copy. In the special homogeneous noise case, we get balanced noise for the data mode and the ancilla mode in each layer. Thus we will call this type  `balanced'. For the $\ell$th layer, from Eq.~\eqref{eq:sigma_L} we have the logical output noise
\begin{equation}
\begin{aligned}
(\sigma_{\rm L}^{(\ell)})^2\simeq& 2\sigma_{\rm GKP}^2+ \frac{4\left((\bar{\sigma}_{\rm L}^{(\ell-1)})^4-4\sigma_{\rm GKP}^4\right)}{\pi}
\\& \times \ln{  \frac{\pi^{\frac{3}{2}}}{2\left((\bar{\sigma}_{\rm L}^{(\ell-1)})^4-4\sigma_{\rm GKP}^4\right)} }.
\label{eq:sigma^n}
\end{aligned}
\end{equation}
Similarly, we can obtain the converged value $\sigma_{\infty}$ via solving Eq.~\eqref{eq:sigma^n} after setting $\bar{\sigma}_{\rm L}^{(\ell)}=\bar{\sigma}_{\rm L}^{(\ell-1)}=\sigma_{\infty}$. 
%\begin{equation}
%\begin{aligned}
%(\sigma_\infty)^2\simeq& 2\sigma_{\rm GKP}^2+ \frac{4\left(\sigma_\infty^4 -4\sigma_{\rm GKP}^4\right)}{\pi}
%\\& \times \ln{  \frac{\pi^{\frac{3}{2}}}{2\left(\sigma_\infty^4-4\sigma_{\rm GKP}^4\right)} },
%\label{eq:sigma^n}
%\end{aligned}
%\end{equation}
The solution can be obtained analytically as
\begin{equation}
\label{eq:LB}
(\sigma_\infty)^2=2\sigma_{\rm GKP}^2,
\end{equation}
which achieves the ultimate lower bound set by the GKP noise.

\begin{figure}
\centering
\includegraphics[width=\linewidth]{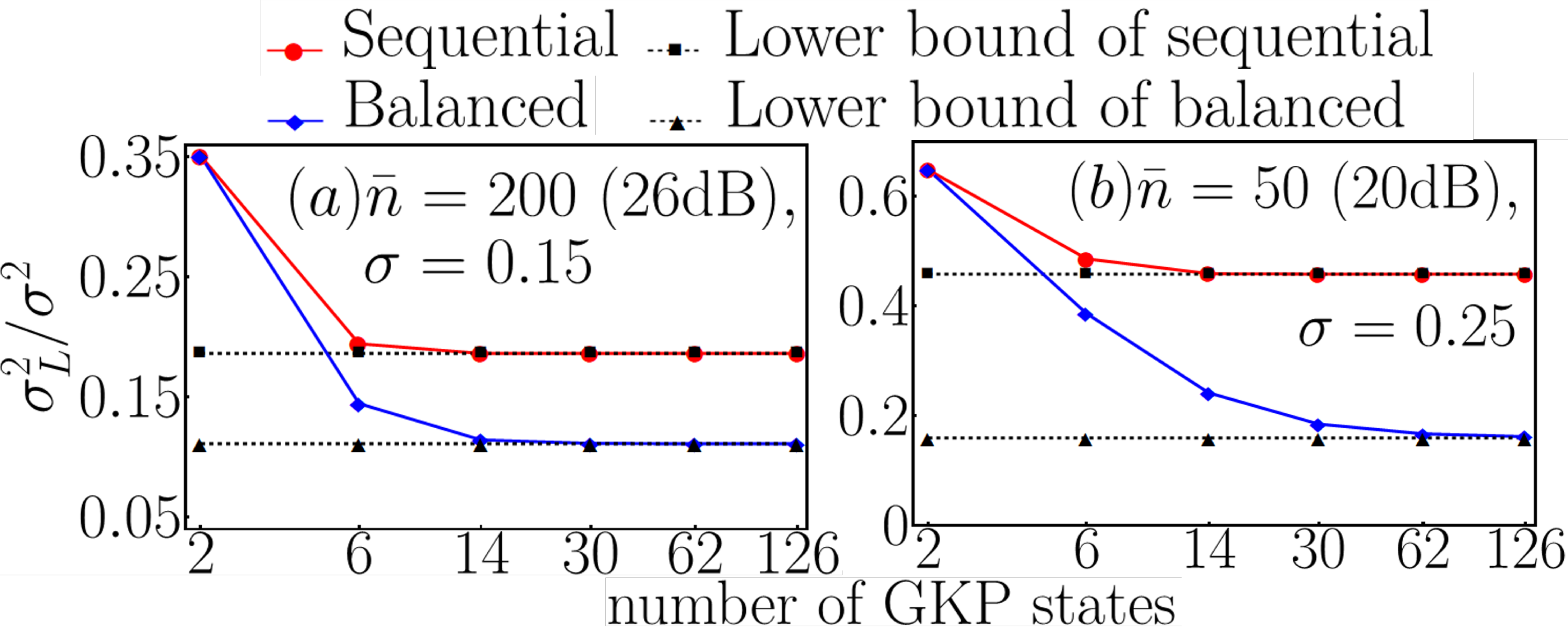}
\caption{Performance of sequential type (Red line), and balanced type (Blue line),\bz{ with (a) initial $\sigma=0.15$ and mean photon number in ancillary modes $\bar{n}=200$ (the GKP squeezing is defined as $s_{\rm GKP}=-10\log_{10}(2\sigma_{\rm GKP}^2)$, corresponding to $26\text{dB}$), and (b) initial $\sigma=0.25$ and $\bar{n}=50$ (corresponding to $20\text{dB}$). Dashed lines are the lower bound of each concatenation codes.df} }
\label{fig.2}
\end{figure}

Now we numerically evaluate the performance of the sequential and balanced types of concatenated codes in Fig.~\ref{fig.2}, where we use the output versus input noise ratio $\sigma_{\rm L}^2/\sigma^2$ as a figure of merit. \bz{The black dashed lines with squares} show the lower bound of noise level $\sigma_{\infty}$ of the sequential type, and \bz{the black dashed lines with triangles} show the lower bound $\sigma_{\infty}$ of balanced type which is also the ultimate lower bound of concatenation codes. Comparing subplots (a) and (b), we see that as the number of GKP states in the encoding increases, the concatenation codes perform better. \bz{Since the code is effective when $\sigma_{\rm GKP}<{\sigma}/{\sqrt{2}}$, we have a constraint on the mean photon number in the GKP ancilla needed for QEC. For instance, using $\sigma_{\rm GKP}^2=1/(4\bar{n})$, we have that $\bar{n}>22$ required when $\sigma=0.15$ (equivalent to $16\text{dB}$ of squeezing in the GKP state); for $\sigma=0.25$, $\bar{n}>8$ (or $12\text{dB}$ of squeezing). We also find that as $\bar{n}$ decreases, the balanced type QEC needs more ancilla GKP states to converge to the ultimate lower bound. This can be seen in Fig.~\ref{fig.2}. In subplot (a), we choose $\bar{n}=200$ ($26\text{dB}$), where fast convergence is observed. In subplot (b), as an example of small mean photon number regime, we choose $\bar{n}=50$, which corresponds to the value of the GKP squeezing $20\text{dB}$. If we use even less $\bar{n}$ (i.e., less GKP squeezing), the number of GKP states required to reach the lower bound becomes impractical.}

Before comparing these concatenation schemes any further, let us point out that each scheme consumes a different number of ancillary GKP states in each layer. For the sequential scheme, each layer will consume 2 ancillary GKP states (one in the encoding procedure, another in the measurement procedure); so $n$ layers will consume $2n$ total ancillary GKP states. For the balanced scheme, the $n$th concatenation layer will consume $2^{n}$ ancillary GKP states; so $n$ layers will consume $\sum_{\ell=1}^n 2^\ell =2(2^{n}-1)$ ancillary states totally. For fair comparison, we take the number of GKP-states consumed as a resource. 

%\begin{figure*}[htbp]
%\begin{center}
%\includegraphics[width=0.9\linewidth]{FIG3_new.png}
%\end{center}
%\caption{Performance of sequential type (Red line), and balanced type (Blue line), with initial %$\sigma=0.15$, (a) mean photon number in ancillary modes $\bar{n}=100$ (corresponding to noise %$\sigma_{\rm GKP}=0.05$), and (b) $\bar{n}=400$ (corresponding to noise $\sigma_{\rm GKP}=0.025$). %Dashed line is the lower bound of the concatenation codes.\bz{example. Then need to add analysis.} }
%\label{fig.2}
%\end{figure*}

Since we consider homogeneous noise $\sigma$, the result of the first layer $(\sigma^{(1)}_{\rm L})^2$ is the same for each concatenation scheme. However, upon increasing the layers, we find that the balanced scheme has a faster convergence than sequential scheme. After several layers of concatenation, each concatenation scheme approaches its respective lower bound $\sigma_\infty$. We note, however, that only the balanced scheme reaches the theoretical minimum, $2\sigma_{\rm GKP}^2$. The reason is due to the fact that the sequential scheme always contacts with the initial (and also the largest) noise $\sigma$ which cannot be lowered by this scheme. Contrariwise, in the balanced scheme, the reduced output always contacts with the same reduced noise, so the effect is always better.

\begin{figure}
\centering
\includegraphics[width=\linewidth]{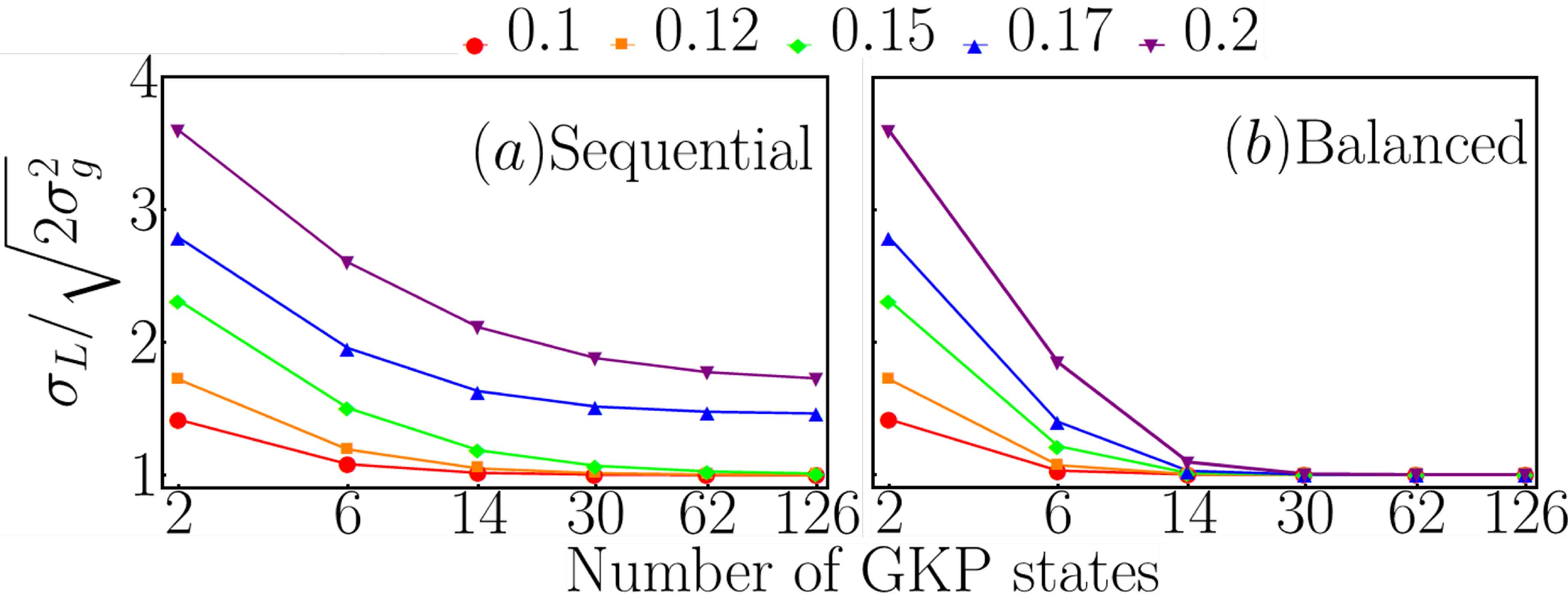}
\caption{Performance of concatenation codes of different initial $\sigma$, with constant mean photon number in ancilla modes $\bar{n}=400$ ($\sigma_{\rm GKP}=0.025$) (a) sequential type, and (b) balanced type. Red lines with circles present $\sigma=0.1$, orange lines with squares present $\sigma=0.12$, green lines with diamonds present $\sigma=0.15$, blue lines with up triangles present $\sigma=0.17$, and purple lines with down triangles present $\sigma=0.2$.} 
\label{fig.3}
\end{figure}

Fig.~\ref{fig.3} shows how different initial input-noises will converge to the lower bound for each concatenation scheme. Here, we assess the convergence with the figure of merit, $\sigma_{\rm L}/\sqrt{2\sigma_{\rm GKP}^2}$, where denominator is the theoretical minimum. Fig.~\ref{fig.3} (a) (the sequential scheme) shows that a larger input noise will correspondingly converge to a larger limiting value, which depends on the initial noise and $\sigma_{\rm GKP}$ as can be observed in Eq.~\eqref{eq:sigma^n_sequential} setting $\bar{\sigma}_{\rm L}^{(\ell)}=\bar{\sigma}_{\rm L}^{(\ell-1)}=\sigma_{\infty}$; however, when the initial $\sigma$ is fairly small, the lower bound approaches the theoretical minimum of $\sqrt{2\sigma_{\rm GKP}^2}$. On the other hand, Fig.~\ref{fig.3} (b) (the balanced scheme) shows that, even when each scenario has drastically different output noise in the first concatenation layer, soon they will all converge to the same theoretical lower bound in 3 or 4 layers of concatenation (14-30 ancillary GKP states), since the lower bound only depends on $\sigma_{\rm GKP}$.

%\bz{Discard this paragraph.}  All other types are the middle or the combination of the sequential type and balanced type. Let's start from the first layer, for the initial input, $\sigma_1$, only has one GKP ancillary state can choose, that is $\sigma_1$. So there are only one output, $\sigma_2$. For the second layer, it has two choice, $\sigma_1$ or $\sigma_2$. So the output has two different values, $\sigma_{3,1}$, and $\sigma_{3,2}$. For the third layer, it can choose all the results of the second layer, and the first layer, $\sigma_1$, $\sigma_2$, $\sigma_{3,1}$, and $\sigma_{3,2}$. For the $n$th layer, it has all the results of last layers' output. We can find, all the choice can been seen as two different cases, balanced choice or unbalanced choice. As shown in Fig.~\ref{fig.3}(e), an example of type (c), it's a combination of the sequential type and balanced type, for the second layer, it is a balanced choice, but then it always choose the same $\sigma_2$, which is an unbalanced choice. So we can call this as combined type. Intuitively speaking, since we know balanced type is always better than sequential type, all other possible combination types should between the two types, this is as the Fig.~\ref{fig.3}(e) shows. This combination type has a lower bound between the sequential type and balanced type.

\section{Application 1: GKP codes and distributed sensing}
\label{IV}

\subsection{Error correction in Distributed Quantum Sensing Protocol}
\label{IV A}

DQS aims to estimate a global parameter of multiple local parameters, such as a weighted average. In a classical distributed sensing scenario, when averaging the measurement outcomes at each sensor node, the root-mean-square (rms) estimation error scales as $1/\sqrt{M}$ (the standard quantum limit), where $M$ is the number of measurements. On the other hand, DQS can beat the standard quantum limit by utilizing shared entanglement between various sensors. In particular, in the bosonic case of estimating weighted average of displacements, DQS protocols utilize continuous-variable multi-partite entanglement to perform better than the standard quantum limit. However, practical imperfections such as loss and additive noise severely limit the performance of typical DQS protocols relying purely on Gaussian resources. It is therefore necessary to add a non-Gaussian ingredient in order to increase the robustness. 

\begin{figure}[t]
\centering
\includegraphics[width=\linewidth]{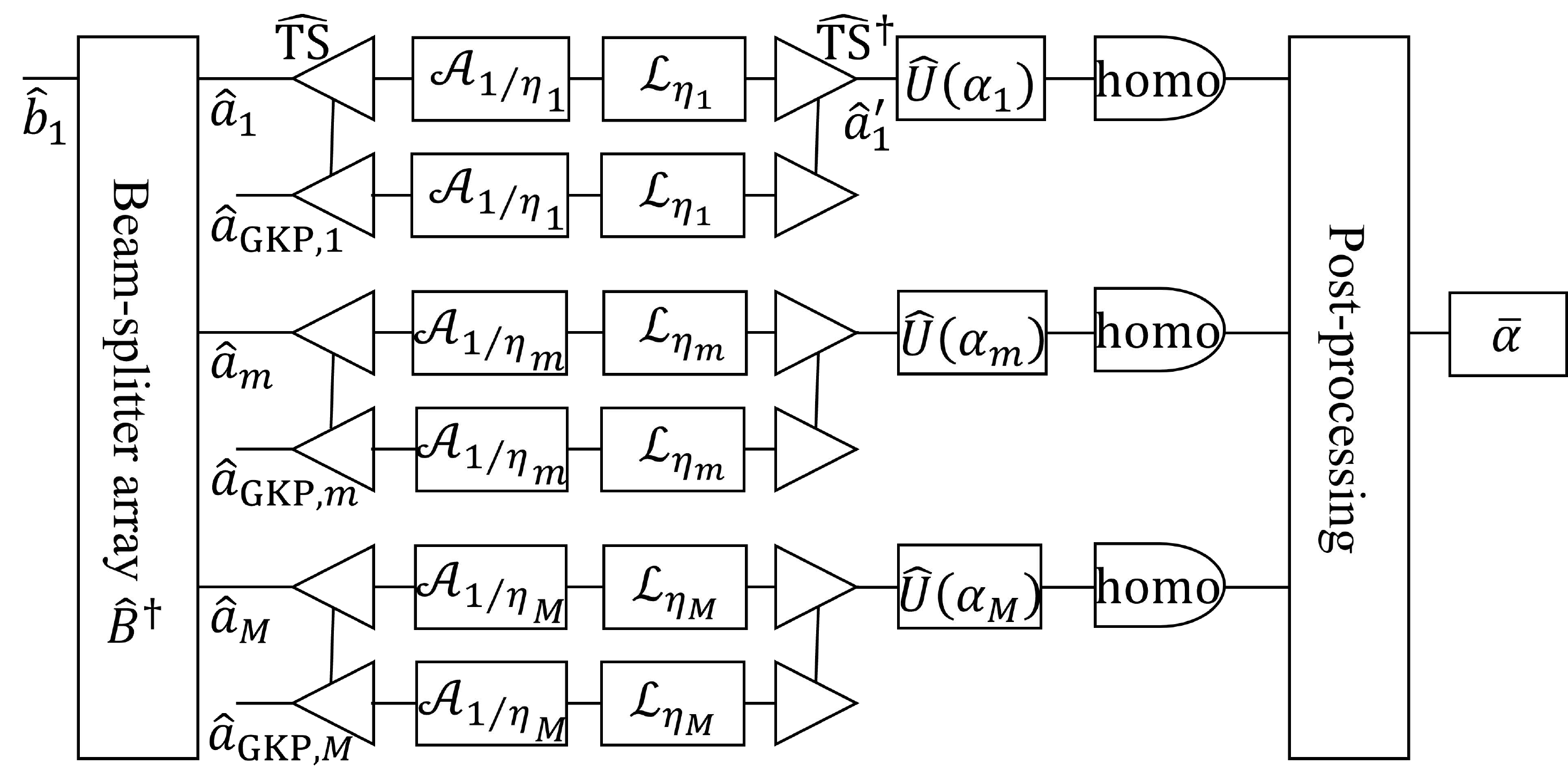}
\caption{Schematic of a distributed quantum sensing protocol with GKP-TMS error correction in the distribution step. The label `$\hat{\rm TS}$' represents two-mode squeezing, while `$\hat{\rm TS}^\dagger$' represents the inverse two-mode squeezing; `homo' represents homodyne detection.}
\label{fig.4}
\end{figure}

Ref.~\cite{zhuang2020distributed} applied GKP-TMS codes to minimize additive Gaussian noise in DQS, which occurs with unit probability on all modes. As shown in Fig.~\ref{fig.4}, the objective is to estimate a weighted average, $\bar{\alpha} \equiv \sum_{m=1}^{M}w_m \alpha_m$, where $w_m$ is the weight of each mode, and $\alpha_m$ is the displacement among each mode. 

%To being with, the DQS protocol mixes a squeezed-vacuum state with $M-1$ vacuum states on a balanced beam-splitter network to produce a multi-partite entangled probe state. Each output modes $\hat{a}_m$ then undergoes a CV-QEC protocol, which mixes a GKP ancilla via the GKP-TMS code. Each mode is then distributed through a noisy channel to the sensor nodes.

To begin with, the multi-partite entangled state of modes \{$\hat{a}_m$, $1\leq m \leq M$\} is produced by passing a single-mode squeezed vacuum \QZ{with mean photon number $N_S$},  through a beam-splitter array $\hat{B}^{\dagger}$. Each output mode $\hat{a}_m$ then undergoes a CV-QEC protocol, which mixes a GKP ancilla via the GKP-TMS code. Each mode is then distributed through a noisy channel to the sensor nodes.

In the encoding circuit, each mode $\hat{a}_m$ is entangled via a two-mode squeezing
operation with an ancilla mode in the GKP state. %In the Heisenberg picture, the annihilation operators are transformed by a linear unitary transformation. 
Following the two-mode squeezing operations on each mode, an amplification channel is applied just before distribution of the modes through a pure-loss channel. The concatenation of the amplification followed by loss converts the imperfections to additive white Gaussian noise (AWGN) (see Appendix~\ref{app:noise_reduction}). Upon reception of the modes at the corresponding receiver stations,  a decoding circuit is applied, which is the inverse of encoding circuit. Note that, during encoding and decoding, the additive Gaussian
noise adds position noise $\xi_q$ and momentum noise $\xi_p$. Finally, a local displacement, $\hat{\bm{U}}(\alpha)$, occurs for each mode, and a homodyne measurement is performed to measure the real quadrature $\mathbf{Re}(\hat{a}_m^{\prime})$. Combining all measurement data after post-processing, one obtains an estimation of $\bar{\alpha}$.

We sum all the homodyne results by the weights to obtain the estimation of $\bar{\alpha}$.  
Let $\bar{w}=\sqrt{\sum_{m=1}^M w_m^2}  $, and \be 
\bar{\sigma}_{\rm L}=\frac{1}{\bar{w}}\sqrt{\sum_{m=1}^M w_m^2 \sigma^2_{\rm{L}_m}},
\label{sigma_L}
\ee 
where $\sigma_{\rm{L}_m}$ is the logical output noise of each senor mode, given by Eq.~\eqref{eq:sigma_real_main} with parameters for each sensor. The rms error of estimating the parameter $\alpha$ can be derived as~\cite{zhuang2020distributed}
\begin{equation}
\begin{aligned}
\delta \alpha =\frac{\bar{w}}{2}\left(\frac{1}{(\sqrt{N_S+1}+\sqrt{N_S})^2}+2\bar{\sigma}_{\rm L}^2 \right)^{\frac{1}{2}},
\label{eq:dalpha}
\end{aligned}
\end{equation}
\QZ{where $N_S$ is the mean photon number of the initial squeezed vacuum.}

\subsection{Performance evaluation}
\label{IV B}

From Eq.~\eqref{eq:sigma_L}, we can easily get a conclusion that the output $\bar{\sigma}_{\rm L}$ decreases as the noisy $\sigma_{\rm GKP}$ decreases, until the noise is zero, in which case the output reduces to the same value as the ideal GKP state. However, this convergence to zero noise requires infinite energy. A more practical way is to consider a reasonable distribution of energy to the whole system. To get the minimal variance $\delta \alpha^2$, we need to find the minimal $\bar{\sigma}_{\rm L}^2$ as specified in Eq.~\eqref{eq:dalpha}. %So the point is to find the best method of energy distribution to get the minimal. 

First, we assume that the total amount of energy (mean photon number) across all GKP ancilla is limited. To find the best method, consider each sensor's GKP noise $\sigma_{\rm GKP,m}^2=1/4\bar{n}_m$ and Eq.~\eqref{eq:sigma_L}, 
where $\bar{n}_m$ is the mean photon number in GKP ancillary states of $m$th sensor, and $w_m$ is the weight of $m$th sensor. While $\sigma_{\rm GKP,m}$ appears in both the first term and the second term in Eq.~\eqref{eq:sigma_L}, the dependence of the first term on $\sigma_{\rm GKP,m}$ is dominant. Consequently, the problem of minimizing the overall noise $\bar{\sigma}_L$ in Eq.~\eqref{sigma_L} is reduced to minimizing $\sum_{m=1}^M w_m^2 (1/2\bar{n}_m)$, under the photon-number constraint $\sum_{m=1}^M \bar{n}_m=\bar{n}$. By using the method of Lagrange multiplier, we formally obtain the asymptotic optimal distribution ratio ${\bar{n}_m}/{w_m}=\mathrm{ Constant}$.
This means the best output strategy is to give each mode the same proportion of photons as their weights.%, even when the noise between sensors differ. 
\begin{figure}[t]
\centering
\includegraphics[width=\linewidth]{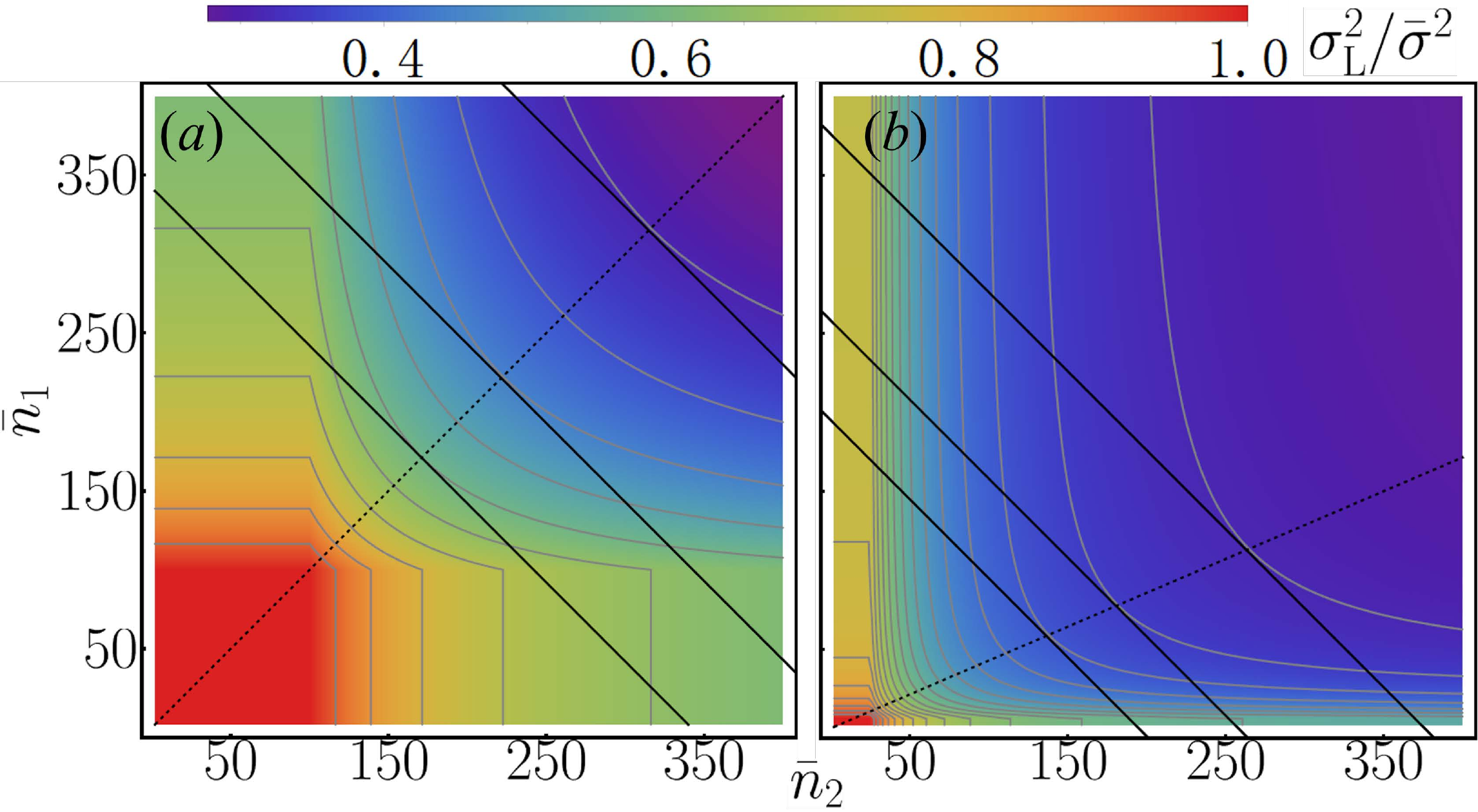}
\caption{Contour plot of the two sensor modes, with (a)identical noise $0.05$ and weight $w_1=w_2=0.5$, and (b)asymmetric noise $0.2, 0.1$, with weight $w_1=0.3,w_2=0.7$. X-axis is the photon numbers of mode 2,  Y-axis is the photon numbers of mode 1. Contours are the ratio of output $\sigma_{\rm L}^2$ with and without error correction.} 
\label{fig.5}
\end{figure}

We verify the above results in an example of two-sensor case as shown in Fig.~\ref{fig.5}. The first example in Fig.~\ref{fig.5} (a) involves identical noise standard deviation $\sigma= 0.05$ and weight $w_1=w_2=0.5$ for sensor 1 and sensor 2, and the second example in Fig.~\ref{fig.5} (b) involves asymmetric noise standard deviations of $\sigma=0.2,0.1$ and weights $w_1=0.3,w_2=0.7$ for sensor 1 and sensor 2. Contour values represent the ratio of $\bar{\sigma}_{\rm L}^2$ with and without error correction process. \bz{The red area shows where $\sigma_{\rm L}^2$ is the same as the input noise $\bar{\sigma}^2$ without error correction process. In Fig.~\ref{fig.5} (a), for small input noise $\sigma=0.05$, we require at leas $\bar{n}_1>100$ and $\bar{n}_2>100$ GKP photons for the error correction code to be effective. For larger input noise, around $\sigma=0.2, 0.1$, the number of GKP photons required for successful error correction relaxes to around $\bar{n}_1>15$ and $\bar{n}_2>40$; see Fig.~\ref{fig.5} (b).} In Fig.~\ref{fig.5} (a), three solid lines represent constant total numbers of $\bar{n}=342,443,630$, and in Fig.~\ref{fig.5} (b), three solid lines represent constant total numbers of $\bar{n}=240,306,425$. Each line has only one tangent point in the contour plot, and all other points along each line are larger than the tangent point. Furthermore, each line's tangent points are also along the dashed line, which represents the initial weight of each sensor, ${\bar{n}_1}/{\bar{n}_2}=1$ in (a) and ${\bar{n}_1}/{\bar{n}_2}={3}/{7}$ in (b), thus verifying our theoretical analyses.

\section{Application 2: Supervised Learning Assisted by an Entangled Sensor Network}
\label{V}

Refs.~\cite{zhuang2019physical,xia2021quantum} introduced a new quantum supervised learning scheme--- supervised learning assisted by an entangled sensor network (SLAEN). SLAEN consists of a hybrid quantum-classical framework, with the core architecture being that of an entangled sensor-network configured by a classical support-vector machine for quantum data classification. The performance enhancement of SLAEN over the classical supervised learning schemes is due to the entanglement shared amongst the different sensors. Entanglement can reduce errors and boost the sensitivity, which has been theoretically proved and experimentally shown. In this context, CV-DQS has the potential to capture global features of an interrogated object better than separable sensors, however in the presence of loss, the performance of SLAEN is restricted. We are thus led to the idea that GKP-TMS quantum error correction codes can assist with SLAEN.

To understand the performance improvement of GKP-TMS error correction code to mitigate loss and noise in the context of SLAEN (and how this advantage generally translates to channel-classification problems), we introduce a simple binary channel discrimination task to calculate the error-probability performance. The hypothesis $H$ that we are testing is a binary random variable with two possible values, $0$ and $1$, whose corresponding prior probabilities, $\pi_0$ and $\pi_1$, are normalized $\pi_0+\pi_1=1$. We map the binary outcome $1$ to the vector $\Vec{\alpha}=(\alpha_1,...,\alpha_M)^{T}$ and the binary outcome $0$ to the vector $\Vec{\beta}=(\beta_1,...,\beta_M)^{T}$ with $\alpha_i,\beta_i\in\mathbb{R}$. Here $M$ is the number of sensor nodes and also the number of probe modes. We consider the channels being classified as quadrature displacement on $M$ sensors, and each vector represents the displacement on the sensors---$\alpha_m$ and $\beta_m$ are the possible displacement amplitudes on the $m$th sensor.

Ref.~\cite{zhuang2019physical} has obtained the optimum Gaussian input state that minimizes the error probability and also introduced a more practical scheme, which is to perform a homodyne detection with a
maximum-likelihood decision rule. In this case, the quantum circuit layout is identical to that in Fig.~\ref{fig.4}, where transmissivities of the distribution channel $\{\eta_m\}$ can be heterogeneous. Let $\pi_0=\pi_1=1/2$, we then have the error probability
\begin{equation}
P_E=\frac{1}{2}\mathrm{Erfc} \left(\frac{||\Vec{\beta}-\Vec{\alpha}||}{2\sqrt{2} \delta }\right).
\end{equation}
where the distance $||\Vec{\beta}-\Vec{\alpha}||$ is a $L^2$ norm, and $ \delta $ is the measurement standard deviation that we specify below. Using the error probability as a figure of merit for SLAEN, we compare four different sensing scenarios:

\begin{enumerate}

\item 
Entanglement assisted distributed sensing scheme in a heterogeneous, noisy environment (with $0<\eta_m<1$, $\forall\,m$) without any error correction. We have the measurement noise
\be 
\delta^{\rm EA}=\frac{\bar{w}}{2}\Bigl( \frac{\bar{\eta}}{(\sqrt{N_s+1}+\sqrt{N_s})^2}+1-\bar{\eta} \Bigr)^{\frac{1}{2}},
\ee 
where $\bar{w}=\sqrt{\sum_{m=1}^M w_m^2}  $, and $\bar{\eta}=\sqrt{\sum_{m=1}^M w_m^2 \eta_{m}}/\bar{w}$. 

When $\eta_m=1$ and $w_m=1/M$ (no loss and equal weights), this reduces to the ideal case, where the optimal input is a squeezed-vacuum state distributed uniformly across all modes, and we denote it as $\delta^I$. For a single-mode ($M=1$), the error probability expression degenerates to Eq. (A5) in Ref.~\cite{zhuang2019physical}.

\item 
Distributed sensing scheme with GKP-TMS error correction code in a heterogenous environment. In Sec.~\ref{IV A}, we analysed the performance of the GKP-TMS error correction code in the face of additive Gaussian noise. The measurement noise $\delta^{\rm{GKP}}$ has the same expression as Eq.~\eqref{eq:dalpha}. (We use $\delta$, as opposed to $\delta\alpha$, to avoid confusion.)

\item 
    Distributed sensing with GKP-TMS concatenation codes. In Sec.~\ref{III B}, we have the lower bounds for the balanced scheme and sequential scheme. So we use the lower bound result for this condition. The measurement noise lower bound $\delta^{\rm LB}$ can be obtained via utilizing $\bar{\sigma}_{\rm L}^{(\infty)}$ to replace $\bar{\sigma}_{\rm L}$ in Eq.~\eqref{eq:dalpha}, where $\bar{\sigma}_{\rm L}^{(\infty)}$ represents the theoretical lower bound after an infinite number of concatenations. For the balanced scheme, $\bar{\sigma}_{\rm L}^{(\infty)}=2\sigma_{\rm GKP}^2$ [see relation~\eqref{eq:LB}]. For the sequential scheme, we approach the lower bound numerically after consuming about 30 GKP states.

\item Optimal separable-state scheme. For comparison, we include the case of $M$ independent sensors, each with a single-mode squeezed vacuum input. The measurement noise $\delta^{\rm S}$ has same expression as Eq.~\eqref{eq:dalpha}, except we need to use the mean photon number $N_m$ in each sensor mode to replace  the total mean photon number $N_s$, where $N_m=N_s/M$, and $\sum_{m=1}^M N_m=N_s$.

\end{enumerate}

\begin{figure}[t]
\centering
\centering
\includegraphics[width=\linewidth]{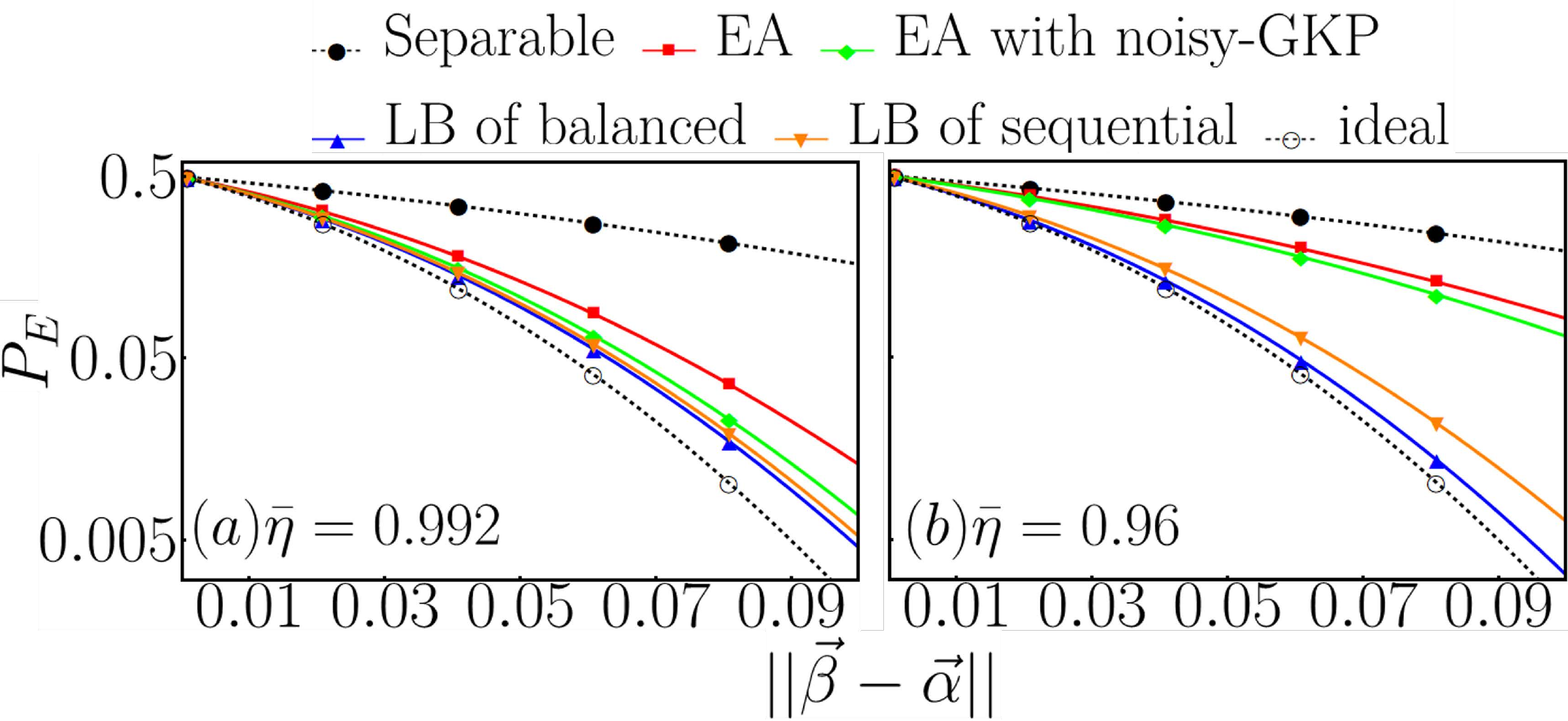}
\caption{ Error probability of different schemes, assuming $N_s=20$ and $M=10$.
\bz{Black dashed lines with filled circles represent the separable state scheme. Red lines with squares represent no-error correction. Green lines with diamonds represent a one-layer finite squeezed GKP error correction code. Orange lines with inverted triangles represent the lower bound of sequential concatenation code. Blue lines with regular triangles represent the lower bound of balanced concatenation code. Black dashed lines with empty circles represent ideal distributed sensing with no loss. Mean photon number in GKP states $\bar{n}=400$, (a) $\eta\in [0.99,1],\bar{\eta}=0.992$, and (b) $\eta\in [0.9,1],\bar{\eta}=0.96$} }
\label{fig.7}
\end{figure}

In Fig.~\ref{fig.7}, we plot the error probability for the each sensing scenario versus the distance $||\Vec{\beta}-\Vec{\alpha}||$. All plots assume $N_s=20$ for the squeezed states and $M=10$ for the number of modes, with mean photon number in GKP states $\bar{n}=400$, (a) transmissivity $\eta\in [0.99,1]$ in each modes, with average $\bar{\eta}=0.992$, and (b) $\eta\in [0.9,1],\bar{\eta}=0.96$. In each case, we consider the direction of $\Vec{\beta}-\Vec{\alpha}$ such that an equal weight is optimal for data classification.

The black dashed lines with empty circles represent the ideal lossless case, which have the smallest error probability since there is no noise. The black dashed lines with filled circles are the performance of the separable scheme in condition (4), which have the largest error probability. The red lines with squares represent condition (1), entanglement-assisted distributed sensing scheme, which have the second largest error probability. This can be lowered by using GKP error correction codes, as conditions (2) (green lines with diamonds) and (3) (blue lines with up triangles and orange lines with down triangles) show. The concatenation codes can go further to lower the noise especially when the additive Gaussian noise parameter $\sigma$, is reasonably large. For instance, Ref.~\cite{noh2020encoding} showed the single layer GKP-TMS code is effective only when $\sigma <0.558$, which translates to an attenuation threshold of $\eta>0.689$, however the benefits of the GKP-TMS code are not completely obvious when $\eta$ is close to this threshold value. On the other hand, concatenation codes can markedly lower the output noise even when $\eta$ is close to threshold. This effect is most stark for multiple concatenation layers because, in this case, one approaches the lower bound of the GKP-TMS code blue lines with up triangles and orange lines with down triangles). In these cases, we see that the sequential scheme is not as effective as the balanced scheme, especially when $\eta$ is relatively large. This is consistent with our conclusions in Sec.~\ref{III B}.

\section{Discussion and Conclusion}
In this paper, we generalize the ideal GKP-TMS error correction codes to finite squeezed GKP-TMS with heterogeneous Gaussian noises, which is more realistic in experiments. While the limit of noise reduction from error correction is determined by the noise $\sigma_{\rm GKP}$ of the finite squeezed GKP ancilla states, different concatenation codes will behave differently towards approaching it. For example, we show that a balanced concatenation outperforms a sequential concatenation, when multiple modes are available to error-correct.

We also applied the finite squeezed GKP-TMS codes to DQS. For estimating a weighted average of displacements under an energy constraint for the ancillary GKP modes, we found that the (asymptotic) optimal strategy to lower the total output noise is to divide the total number photons by the weights at each sensor. We further applied the GKP-TMS code to a channel classification problem. All the schemes with finite squeezed GKP-TMS codes can lower the error probability of entanglement assisted distributed sensing, with respect to no error correction. This is especially true for the balanced concatenation scheme, as it is the closest to the ideal scheme with no noise.

Finally, we point out a future direction for concatenation codes. We have compared two basic concatenation schemes, and the performance of all other constructions based on these two basic types should fall somewhere in between them. However, it is an open problem whether there is another concatenation scheme that approaches the ultimate lower bound set by the balanced scheme but with a faster convergence rate, and while consuming fewer ancillary GKP states.

\begin{acknowledgments}
This work is supported by the Defense Advanced Research Projects Agency (DARPA) under Young Faculty Award (YFA) Grant No. N660012014029. Q.Z. also acknowledges support from NSF OIA-2134830 and NSF OIA-2040575.
\end{acknowledgments}

\appendix

\section{\label{appendix A} Detailed derivation the logical noise}

Combining the noise measurement result of Eq.~\eqref{final_estimator} and final noise of Eq.~\eqref{final_noise}, we can derive the logical noise variance
\begin{widetext}
\begin{equation}
\begin{aligned}
\sigma_{\rm L}^2 &=\sum_{n \in \mathbb{Z} }\int_{\mathbb{R}^2} d\xi_q^{\rm GKP}dz_q^{(2)}p\left[\sqrt{s}\right]\left(z_q^{(2)}\right)
 \cdot p\left[\sqrt{2}\sigma_{\rm GKP}\right]\left(\xi_q^{\rm GKP}\right) \times
 \\&
 \Biggl\{\frac{\sigma_1^2 \sigma_2^2}{s}
   + \Bigl[\frac{t}{s+2\sigma_{\rm GKP}^2}\left(\xi_q^{\rm GKP}-\sqrt{2\pi}n\right) 
  -\frac{2t\sigma_{\rm GKP}^2}{s\left(s+2\sigma_{\rm GKP}^2\right)}z_q^{(2)}\Bigr]^2 \Biggr\}
 \times \mathbf{I}\{ z^{(2)}_q +\xi_q^{\rm GKP}\in [(n-1/2)\sqrt{2\pi},(n+1/2)\sqrt{2\pi}] \},
\end{aligned}
\label{eq:sigma_real}
\end{equation}
\end{widetext}
where $\mathbf{I}(C)$ is an indicator function, i.e., $\mathbf{I}(C)=1$ if $C$ is true, or $\mathbf{I}(C)=0$ if $C$ is false; and $p[\sigma](\cdot)$ is the probability density function of a zero-mean Gaussian distribution with standard deviation $\sigma$. To remind ourselves, we have defined $t=\sqrt{G(G-1)}(\sigma_1^2+\sigma_2^2)$ and $s=(G-1)\sigma_1^2+G\sigma_2^2$ in the main text.

To further simplify the notation, we denote the ancilla measurement position noise in the second mode as $x=z_q^{(2)}$ and the additional measurement position noise due to the noisy GKP states $y=\xi_q^{\rm GKP}$. Also let $P_x=p[\sqrt{s}](z_q^{(2)})$, $P_y=p[\sqrt{2}\sigma_{\rm GKP}](\xi_q^{\rm GKP})$, and 
\begin{equation}
\begin{aligned}
&a
=\frac{t}{s+2\sigma_{\rm GKP}^2},
b
=\frac{2t\sigma_{\rm GKP}^2}{s(s+\sigma_{\rm GKP}^2)}.
\end{aligned}
\end{equation}

With the above notation, Eq.~\eqref{eq:sigma_real} can be simplified,
\begin{widetext}
\begin{align}
\sigma_{\rm L}^2 &=\sum_{n \in \mathbb{Z} }\int_{\mathbb{R}^2} dxdy  P_x P_y\times
  \Biggl(\frac{\sigma_1^2 \sigma_2^2}{s}+\left(a(y-\sqrt{2\pi}n)-bx\right)^2\Biggr)
  \times \mathbf{I}\{x +y\in [(n-1/2)\sqrt{2\pi},(n+1/2)\sqrt{2\pi}] \}
 \\& =\sum_{n \in \mathbb{Z} }\int_{\mathbb{R}^2} dxdy \cdot P_x P_y\times
   \Biggl(\frac{\sigma_1^2 \sigma_2^2}{s} + a^2y^2+2\pi a^2n^2 + b^2x^2 
  -\sqrt{2\pi}a^2ny -\sqrt{2\pi}b^2nx - 2abxy\Biggr)
  \nonumber
  \\& \quad \quad \quad \quad \quad \quad \quad \quad \times \mathbf{I}\{x +y\in [(n-1/2)\sqrt{2\pi},(n+1/2)\sqrt{2\pi}] \}
  \\
  &=\frac{\sigma_1^2 \sigma_2^2}{s}+\frac{2t \sigma_{\rm GKP}^2}{s+2\sigma_{\rm GKP}^2}+\sum_{n \in \mathbb{Z} }2\pi a^2 n^2 \int_{\mathbb{R}^2} dxdy  P_x P_y \mathbf{I}\{x +y\in [(n-1/2)\sqrt{2\pi},(n+1/2)\sqrt{2\pi}] \}
    \\
  &=\frac{\sigma_1^2 \sigma_2^2}{s}+\frac{2t \sigma_{\rm GKP}^2}{s+2\sigma_{\rm GKP}^2}+2\sum_{n=1}^\infty 2\pi a^2 n^2 \int_{-\infty}^\infty dx P_x \left[-\mathrm{Erfc}\left(\left(n+1/2\right)\sqrt{2\pi}-x\right)+\mathrm{Erfc}\left(\left(n-1/2\right)\sqrt{2\pi}-x\right)\right]/2
  \label{sigma_L_step}
  \\&=\frac{\sigma_1^2 \sigma_2^2}{s}+\frac{2t \sigma_{\rm GKP}^2}{s+2\sigma_{\rm GKP}^2}+\sum_{n=1}^{n=\infty}2\pi a^2(2n-1)\times \mathrm{Erfc}\left(\frac{(2n-1)\sqrt{\pi}}{2\sqrt{s+2\sigma_{\rm GKP}^2}}\right),
  \label{final_sigma_L}
\end{align}
\end{widetext}
where in Eq.~\eqref{sigma_L_step} we note that the `$\pm n$' terms in the summation are equal and we have integrated out $y$. In the last step of Eq.~\eqref{final_sigma_L}, we have integrated out $x$ via the equation \begin{equation}
\int_{-\infty}^\infty \mathrm{Erfc}(ax+b)\frac{1}{\sqrt{2\pi\sigma^2}}e^{-\frac{(x-\mu)^2}{2\sigma^2}}dx=\mathrm{Erfc}(\frac{a\mu+b}{\sqrt{1+2a^2\sigma^2}})
\end{equation} and combined terms from $n$ and $n-1$ summation.

To obtain asymptotic results, we only keep the $n=1$ term in Eq.~\eqref{final_sigma_L} and utilize
$\lim_{x \to \infty}\mathrm{Erfc}(x) = e^{-x^2}/x\sqrt{\pi}$. We mention two different approximations in the main text, $\sigma_1\approx\sigma_2$ and $\sigma_1 \ll \sigma_2$.
Under the first condition $\sigma_1\simeq \sigma_2$, we have $t^2 \approx s^2$. 
Let $w={1}/{(s+2\sigma_{\rm GKP}^2)}$, we arrive at
\begin{equation}
\begin{aligned}
   \sigma_{\rm L}^2\simeq f(w)\equiv & \frac{\sigma_1^2 \sigma_2^2}{\frac{1}{w}-2\sigma_{\rm GKP}^2}+2(\frac{1}{w}-2\sigma_{\rm GKP}^2) \sigma_{\rm GKP}^2 w +
   \nonumber
   \\
   &
 4(\frac{1}{w}-2\sigma_{\rm GKP}^2)^2 w^{\frac{3}{2}} e^{-\frac{\pi w}{4}} .
 \label{eq:f(w)}
\end{aligned}
\end{equation}
The optimum $w^*$ can be found by solving $f'(w^*)=0$. 
%Note that $f'(w)$ is given by:
%\begin{equation}
%\begin{aligned}
%f'(w)&=-4\sigma_{\rm GKP}^4+\frac{\sigma_1^2 \sigma_2^2}{\left(2\sigma_{\rm GKP}^2 w -1\right)^2}-\frac{2e^{-\frac{\pi}{4}w}}{w^{\frac{3}{2}}}\left(2\sigma_{\rm GKP}^2 w -1\right)\left(-1-\frac{\pi}{2}w-6\sigma_{\rm GKP}^2w+\pi \sigma_{\rm GKP}^2 w^2\right).
%\end{aligned}
%\end{equation}
Through simple calculations, we can obtain $w^*$ as
\begin{equation}
\begin{aligned}
w^{*}
%&=\frac{4}{\pi}\ln{ \frac{2\left(2\sigma_{\rm GKP}^2 w -1\right)\left(-1-\frac{\pi}{2}w-6\sigma_{\rm GKP}^2w+\pi \sigma_{\rm GKP}^2 w^2\right)}{w^{\frac{3}{2}}\left(-4\sigma_{\rm GKP}^4+\frac{\sigma_1^2 \sigma_2^2}{(2\sigma_{\rm GKP}^2 w -1)^2}\right) }  }
\approx \frac{4}{\pi}\ln{ \frac{\pi }{w^{\frac{1}{2}}\left(-4\sigma_{\rm GKP}^4+\sigma_1^2 \sigma_2^2\right)}  } \approx \frac{4}{\pi} \ln{ \frac{\pi^{\frac{3}{2}}}{2\left(\bar{\sigma}^4-4\sigma_{\rm GKP}^4\right)} },
\end{aligned}
\end{equation}
where $\bar{\sigma}=\sqrt{\sigma_1 \sigma_2}$. Since $\sigma_1$, $\sigma_2$ and $\sigma_{\rm GKP}$ are far smaller than 1, we discard quartic power and square power terms in the calculation. And in the last step, we put $w={4}/{\pi}$ into the expression. Then we can put this result in to Eq.~\eqref{eq:f(w)} to get the optimal value:
\begin{equation}
\begin{aligned}
\sigma_{\rm L}^2&=f(w^*) 
%\approx \bar{\sigma}^4 w^*+2\sigma_{\rm GKP}^2 -4\sigma_{\rm GKP}^4 w^* + 4\left(\frac{1}{w^*}-2\sigma_{\rm GKP}^2\right)^2 {w^*}^{\frac{3}{2}} e^{-\frac{\pi }{4}w^*}
\nonumber
\\
&\approx 2\sigma_{\rm GKP}^2+ \frac{4\left(\bar{\sigma}^4-4\sigma_{\rm GKP}^4\right)}{\pi}\ln{  \frac{\pi^{\frac{3}{2}}}{2\left(\bar{\sigma}^4-4\sigma_{\rm GKP}^4\right)} }.
\end{aligned}
\end{equation}

This is the proof of Eq.~\eqref{eq:sigma_L} in Sec.~\ref{III A}. When $\sigma_1=\sigma_2=\sigma$, this degenerates to independent and identically distributed additive Gaussian noise, as considered in Ref.~\cite{noh2020encoding}.

At the limit of $\sigma_1 \ll \sigma_2$, $(G-1)\sigma_1^2+G\sigma_2^2 \approx G\sigma_2^2$, $t \approx \sqrt{G(G-1)}\sigma_2^2$. We have
\begin{align}
\sigma_{\rm L}^2\simeq 2\sigma_{\rm GKP}^2+ &\frac{4\left(\bar{\sigma}^4 -2\sigma_2^2 \sigma_{\rm GKP}^2 -4\sigma_{\rm GKP}^4\right)}{\pi}
\nonumber
\\
&\times\ln{  \frac{\pi^{\frac{3}{2}}}{2\left(\bar{\sigma}^4 -2\sigma_2^2 \sigma_{\rm GKP}^2 -4\sigma_{\rm GKP}^4\right)} }.
\label{sigma1small_app}
\end{align}

Although practically not relevant, for completeness, we also consider the case when the first mode has larger nose, $\sigma_1 \gg \sigma_2$, $(G-1)\sigma_2^2+G\sigma_1^2 \approx (G-1)\sigma_2^2$. In this case, we have $t \approx \sqrt{G(G-1)}\sigma_2^2$ and
\begin{align}
\sigma_{\rm L}^2\simeq  2\sigma_{\rm GKP}^2+ &\frac{4\left(\bar{\sigma}^4 +2\sigma_2^2 \sigma_{\rm GKP}^2 -4\sigma_{\rm GKP}^4\right)}{\pi}
\nonumber
\\
&\times
\ln{  \frac{\pi^{\frac{3}{2}}}{2\left(\bar{\sigma}^4 +2\sigma_2^2 \sigma_{\rm GKP}^2 -4\sigma_{\rm GKP}^4\right)} }.
\label{sigma2small}
\end{align}

\begin{figure*}[htbp]
\centering
\includegraphics[width=0.8\linewidth]{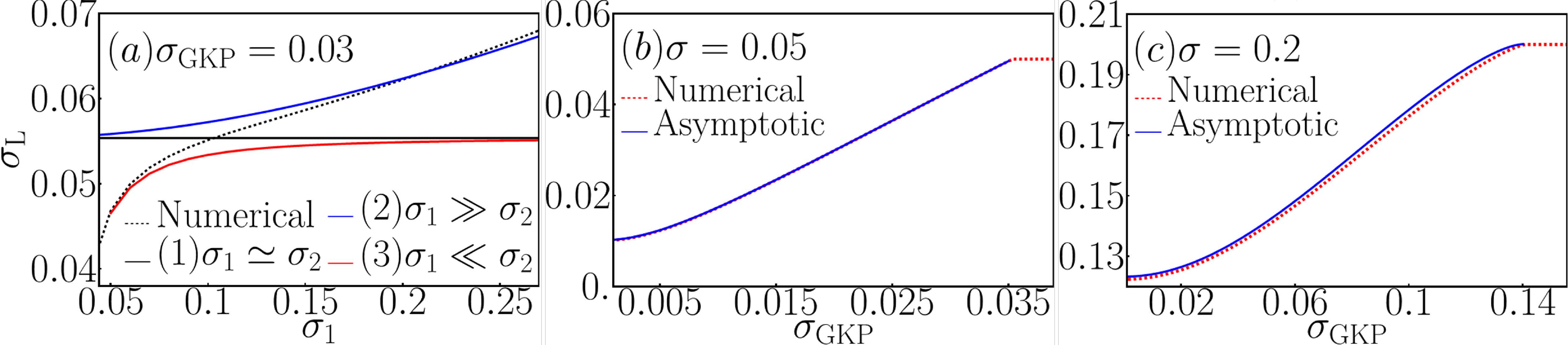}
 \caption{(a) Asymptotic expression for heterogeneous noise. X-axis is $\sigma_1$, and Y-axis is the output logical noise $\sigma_{\rm L}$. In the procedure, we keep $\sigma_1 \times \sigma_2 =\bar{\sigma}^2$ constant and vary $\sigma_1$. (b) and (c) Numerical expression and asymptotic expression result for different $\sigma$. X-axis is the noise of finite squeezed GKP state $\sigma_{\rm GKP}$, Y-axis is the output noise $\sigma_{\rm L}$.}
\label{fig.8}
\end{figure*}

In Fig.~\ref{fig.8} (a), we compare the three asymptotic results with the numerical results, setting $\sigma_1 \times \sigma_2 =\bar{\sigma}^2$ as a constant value. Black solid line represents the asymptotic result $\sigma_{\rm L}$ from Eq.~\eqref{eq:sigma_L} when $\sigma_1\simeq \sigma_2$, which only depends on $\bar{\sigma}$ and is therefore a constant. And when $\sigma_1 \approx \sigma_2$, numerical result (dashed line) is very close to the black solid line. However if one of the input is far larger than another, the other asymptotic expressions (blue for Eq.~\eqref{sigma2small} and red for Eq.~\eqref{sigma1small_app}) are closer to the numerical result. But we need to remember that all these conditions have a  same requirement, that is $G \gg 1$, which can be satisfied when $\sigma_1, \sigma_2 \ll 0.558$.

In Fig.~\ref{fig.8} (b) and (c), we compare the asymptotic results and the numerical results in independent and identically distributed additive Gaussian noise errors ($\sigma_1=\sigma_2=\sigma$) with (b) $\sigma=0.05$, and (c) $\sigma=0.2$. The asymptotic expressions agree pretty well with the exact numerical results, especially when $\sigma$ is in the small regime. And from this figure we can find that $\sigma_{\rm GKP}$ does not affect the fitting accuracy. However,  when $\sigma_{\rm GKP} >{\sigma}/{\sqrt{2}}$, the output $\sigma_{\rm L}$ is the same as the input $\sigma$, and the best $G$ is trivially $G=1$, which also means this code will not help to reduce the noise. Since in this regime, the noise of GKP state is larger than or can compare with the input noise of $\sigma$, it is obvious that this code will fail. From the asymptotic expression, we can also find the regime that the code is effective is $\sigma_{\rm GKP} < {\sigma}/{\sqrt{2}}$, because it will give an imaginary number when $\sigma_{\rm GKP} \geq {\sigma}/{\sqrt{2}}$.

%\begin{figure}
%\centering
%\includegraphics[width=\linewidth]{fig8.pdf}
% \caption{(a) Asymptotic expression for heterogeneous noise. X-axis is $\sigma_1$, and Y-axis is the output logical noise $\sigma_{\rm L}$. In the procedure, we keep $\sigma_1 \times \sigma_2 =\bar{\sigma}^2$ constant %and vary $\sigma_1$. (b) and (c) Numerical expression and asymptotic expression result for different $\sigma$. X-axis is the noise of finite squeezed GKP state $\sigma_{\rm GKP}$, Y-axis is the output noise %$\sigma_{\rm L}$.}
%\label{fig.8}
%\end{figure}

\section{Channel noise reduction} 
\label{app:noise_reduction} 

The Gaussian noise errors accrued in the distribution process are generic in the sense that any excitation-loss error can be converted into an AWGN, when combined with a suitable amplification channel. As an example, consider the excitation loss channel, $\mathcal{L}_\eta$, leading to the mode transformation, 
\begin{equation}
\hat{a}^{\prime}=\sqrt{\eta}\hat{a}+\sqrt{1-\eta} \hat{e},
\end{equation}
where $\hat{e}$ is the vacuum environment mode and $\hat{a}$, $\hat{a}'$ are the input and output modes. Also consider the quantum-limited amplification channel $\mathcal{A}_G$, where
\begin{equation}
\hat{a}^{\prime}=\sqrt{G}\hat{a}-\sqrt{G-1}\hat{e}^{\dagger}.
\end{equation}
By applying a suitable amplification channel, $\mathcal{A}_{1/\eta}$, prior to the loss channel $\mathcal{L}_\eta$, one obtains
\begin{equation}
\begin{aligned}
\hat{a}^{\prime} &=\sqrt{\eta}(\sqrt{1/\eta}\hat{a}-\sqrt{1/\eta-1}\hat{e}^{\dagger})+\sqrt{1-\eta} \hat{e}
\\&=\hat{a}+\sqrt{1-\eta}(\hat{e}-\hat{e}^{\dagger}).
\end{aligned}
\end{equation}
This is just an AWGN channel, $\Phi_{\sigma}$, where $\sigma$ is the variance of the underlying Gaussian distribution (physically $\propto$ the number of quanta added to the mode). In this example,
$\mathcal{L}_\eta \circ \mathcal{A}_{1/\eta}=\Phi_{1-\eta}$.

\section{\label{appendix D} Proof of the lower bound}
In this appendix, we will prove the lower bound of Eq.~\eqref{eq:LB}.
We start from Eq.~\eqref{eq:sigma_real_main}. Since the third term is always positive, we have:
\begin{equation}
\label{eq:lb_calculation}
    \begin{aligned}
    \sigma_{\rm L}^2 & \ge \frac{\sigma_1^2 \sigma_2^2}{s}+\frac{2t \sigma_{\rm GKP}^2}{2\sigma_{\rm GKP}^2+s}
     \\&= \frac{1}{\frac{G-1}{\sigma_2^2}+\frac{G}{\sigma_1^2}}+\frac{2\sqrt{G(G-1)}\sigma_{\rm GKP}^2}{\frac{2\sigma_{\rm GKP}^2}{\sigma_1^2+\sigma_2^2}+\frac{(G-1)\sigma_1^2+G\sigma_2^2}{\sigma_1^2+\sigma_2^2}}.
    \end{aligned}
\end{equation}
To prove the lower bound, we consider the minimization of the R.H.S. of Ineq.~\eqref{eq:lb_calculation}. This depends on the values of $\sigma_1$ and $\sigma_2$, so we will consider two conditions.

When both $\sigma_1,\sigma_2\ge \sqrt{2} \sigma_{\rm GKP}$, Ineq.~\eqref{eq:lb_calculation} can be further lower bounded as 
\begin{equation}
\label{eq:lb_simplify}
    \begin{aligned}
    \sigma_{\rm L}^2 & \ge  \frac{1}{G\left(\frac{1}{\sigma_1^2}+\frac{1}{\sigma_2^2}\right)}+\frac{2\sqrt{G(G-1)}\sigma_{\rm GKP}^2}{\frac{2\sigma_{\rm GKP}^2}{\sigma_1^2+\sigma_2^2}+G}
    \\ &\ge \frac{ \sigma_{\rm GKP}^2}{G}+\frac{4\sqrt{G(G-1)}\sigma_{\rm GKP}^2}{2G+1}.
    \end{aligned}
\end{equation}
This is a monotonically increasing function of $G$, resulting in a tight lower bound with a limiting value of $2\sigma_{\rm GKP}^2$ as  $G \rightarrow \infty$. This is the lower bound of balanced type in the Fig.~\ref{fig.2}.

When $\sigma_1 < \sqrt{2} \sigma_{\rm GKP}$ and $\sigma_1\le \sigma_2$, we need a slightly different approach. Let $x={\sigma_2^2}/{\sigma_1^2}$, then we can rewrite Eq.~\eqref{eq:lb_calculation} as:
\begin{equation}
    \frac{\sigma_{\rm L}^2}{\sigma_1^2} \ge \frac{x}{G-1+Gx}+\frac{2\sqrt{G(G-1)}( 1+x)\frac{\sigma_{\rm GKP}^2}{\sigma_1^2}}{G-1+Gx+2\frac{\sigma_{\rm GKP}^2}{\sigma_1^2}}.
\end{equation}
Since ${\sigma_{\rm GKP}^2}/{\sigma_1^2}>1/2$ and $G>1$, the right hand equation is monotonically increasing with $x$. This is reasonable, because as $\sigma_2$ increases and $\sigma_1$ is held constant, the logical output $\sigma_{\rm L}$ also has to increase. The minimum is when $x=1$, leading to the inequality
\begin{equation}
    \begin{aligned}
    \frac{\sigma_{\rm L}^2}{\sigma_1^2} &\ge \frac{1}{2G-1}+\frac{4\sqrt{G(G-1)}{\sigma_{\rm GKP}^2}/{\sigma_1^2}}{2G-1+2{\sigma_{\rm GKP}^2}/{\sigma_1^2}}
    \\ &=\frac{1}{2G-1}+\frac{4\sqrt{G(G-1)}}{(2G-1){\sigma_1^2}/{\sigma_{\rm GKP}^2}+2}.
    \end{aligned}
\end{equation}
Then, since ${\sigma_1^2}/{\sigma_{\rm GKP}^2}<2$, we have
\begin{equation}
    \begin{aligned}
    \frac{\sigma_{\rm L}^2}{\sigma_1^2}  &> \frac{1}{2G-1}+\frac{4\sqrt{G(G-1)}}{2(2G-1)+2}
    \\ &=\frac{1}{2G-1}+\sqrt{\frac{G-1}{G}}.
    \end{aligned}
\end{equation}
This function has two minima at $G=1$, and $G\rightarrow \infty$, with a minimum value of 1 at each point. Thus, when $\sigma_1 < \sqrt{2} \sigma_{\rm GKP}$ and $\sigma_1\le \sigma_2$, the logical output $\sigma_{\rm L}^2$ has a lower bound $\sigma_1^2$. This means when the noise in the ancilla GKP state is larger than the noise in the data mode, the GKP error correction procedure cannot reduce the noise in the data mode.

%\bibliography{appendix}% Produces the bibliography via BibTeX.
%apsrev4-2.bst 2019-01-14 (MD) hand-edited version of apsrev4-1.bst
%Control: key (0)
%Control: author (8) initials jnrlst
%Control: editor formatted (1) identically to author
%Control: production of article title (0) allowed
%Control: page (0) single
%Control: year (1) truncated
%Control: production of eprint (0) enabled
%

\end{document}